\documentclass[10pt,letterpaper]{article}
\usepackage[top=0.85in,left=2.75in,footskip=0.75in]{geometry}

\usepackage{amsmath,amssymb}

\usepackage{changepage}

\usepackage[utf8x]{inputenc}

\usepackage{textcomp,marvosym}

\usepackage{cite}

\usepackage{nameref,hyperref}

\usepackage{microtype}
\DisableLigatures[f]{encoding = *, family = * }

\usepackage[table]{xcolor}

\usepackage{array}

\newcolumntype{+}{!{\vrule width 2pt}}

\newlength\savedwidth

\newcommand\thickhline{\noalign{\global\savedwidth\arrayrulewidth\global\arrayrulewidth 2pt}%
\hline
\noalign{\global\arrayrulewidth\savedwidth}}

\raggedright
\usepackage[aboveskip=1pt,labelfont=bf,labelsep=period,justification=raggedright,singlelinecheck=off]{caption}

\makeatletter
\renewcommand{\@biblabel}[1]{\quad#1.}
\makeatother

\usepackage{lastpage,fancyhdr,graphicx}
\usepackage{epstopdf}
\fancyheadoffset[L]{2.25in}
\fancyfootoffset[L]{2.25in}
\newcommand{\lorem}{{\bf LOREM}}
\newcommand{\ipsum}{{\bf IPSUM}}

\begin{document}
\vspace*{0.2in}

\begin{flushleft}
{\Large
\textbf\newline{Park factor estimation improvement using pairwise comparison method} 
}
\newline
\\
Eiji Konaka\textsuperscript{1*},
\\
\bigskip
\textbf{1} Dept. of  Information Engineering, Meijo University, Nagoya, JAPAN
\\
\bigskip

%
%





* konaka@meijo-u.ac.jp

\end{flushleft}
\section*{Abstract}
Each ballpark has a different size in baseball. It could be easily imagined that there would be many home runs in a small ballpark.
Moreover, the environment of the ballpark, such as altitude, humidity, air pressure, and wind strength, affects the trajectory of batted balls.

Park Factors (PF) are introduced in baseball to quantify the effect of each ballpark on the results (e.g., home runs).
In this paper, I assume that each plate appearance can be modeled as a match-up between a batter's team and a pitcher's team plus a ballpark.
The effects of each ballpark will be distilled by using a logistic regression method.

Numerical verification shows that the proposed method performs better than conventional PF. The verification is based on the results of more than 1.5 million plate appearances from the 2010 to 2017 Major League Baseball (MLB) seasons.

%


\section*{Introduction}
Each ballpark has a different size in baseball.
It is a unique feature of this sport.
In most other professional sports (e.g., basketball, football, hockey, tennis) the size of the court or pitch is clearly defined.

In baseball, the arrival point and time of a batted ball are important.
Therefore, it is logical to assume that there would be many home runs in a small ballpark.
Moreover, the environment of the ballpark (e.g., altitude, humidity, air pressure, and wind strength) affects the trajectory of batted balls\cite{doi:10.1119/1.2805242,Bahill2009}.

Coors Field, located in Denver, is used as a home ballpark of the Colorado Rockies, a Major League Baseball (MLB) team in the USA.
This ballpark is famous for its high altitude of 1609m above sea level.
The Rockies states on its website: ``It is estimated that a home run hit 400 feet in sea-level Yankee Stadium would travel about 408 feet in Atlanta and as far as 440 feet in the Mile High City.''
\cite{http://colorado.rockies.mlb.com/col/ballpark/history.jsp}

\begin{figure}[h]
\begin{center}
	\includegraphics[width=0.8\columnwidth]{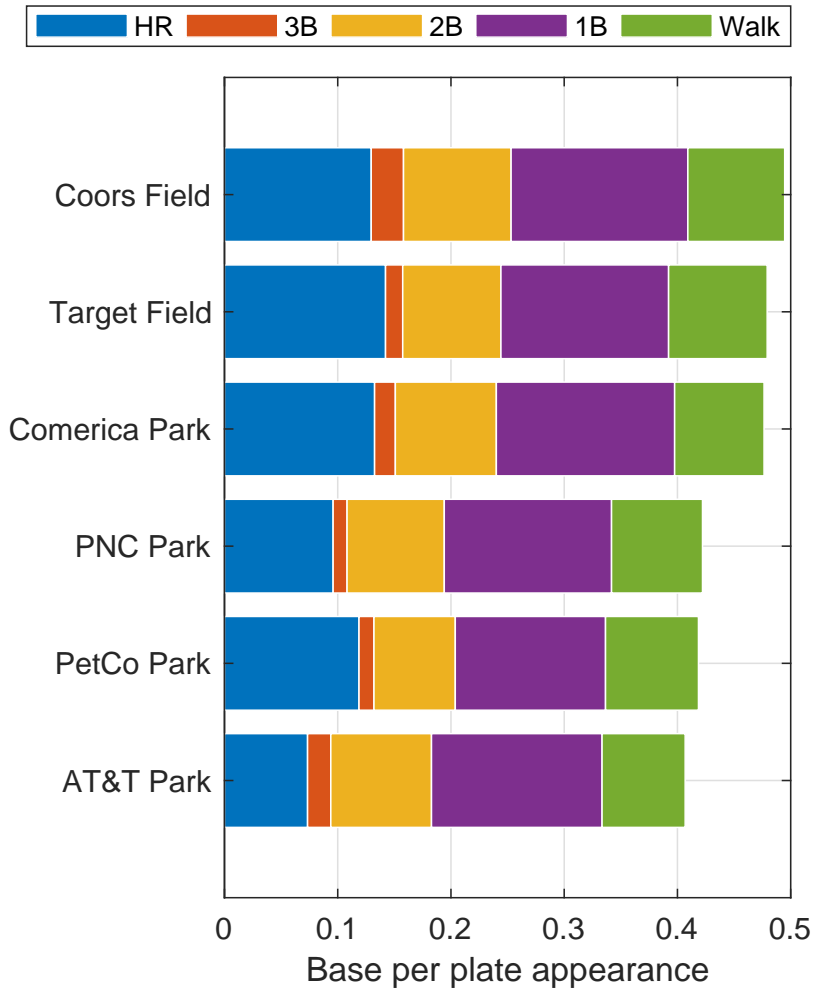}
	\vspace{10pt}
	\caption{Total bases plus walks per plate appearances: top and bottom three (MLB 2017)}
	\label{fig:basesPerPA}
\end{center}
\end{figure}

A fast batted ball with small deceleration is hard for fielders to catch.
Coors Field was ranked first in total bases plus walks per hitter plate appearance (PA) in the 2017 season.
Figure \ref{fig:basesPerPA} shows the top and bottom three ballparks for total bases of walks and hits per hitter PA. The maximum and the minimum values are $0.495$ and $0.4068$, respectively.

\begin{figure}[h]
\begin{center}
	\includegraphics[width=0.8\columnwidth]{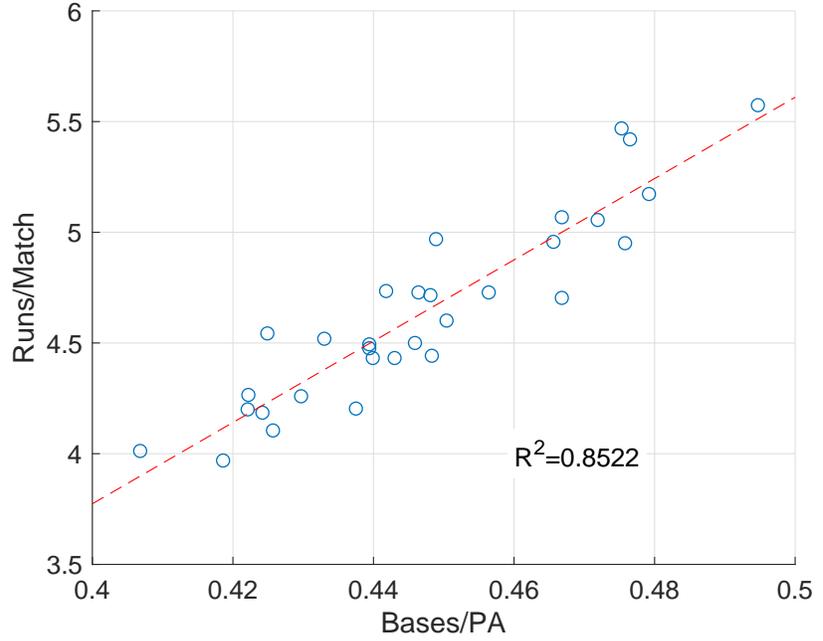}
	\vspace{10pt}
	\caption{Total bases plus walks per plate appearances and runs per match (MLB 2017)}
	\label{fig:basesPerPA_To_runsPerMatch}
\end{center}
\end{figure}

Naturally, bases per PA affects total runs.
Figure \ref{fig:basesPerPA_To_runsPerMatch} illustrates the relationship between bases per PA and total runs per match. Runs are the sum of home and visitor teams.
There are strong correlations between two values with $R^2=0.8522$.

In MLB, every team plays half of the matches in its home park.
Note that the number of matches against other teams differs.
Therefore, the difference of the average bases per PA (Figure \ref{fig:basesPerPA}) could contain both the effects from the players and the park.

Park Factor (PF) has been proposed to quantify each ballpark's effect.
A park with a PF of more than $1.0$ is considered a ``batter-favored'' park, and vice versa.
For instance, ESPN discloses their PF definition and values for every park\cite{http://www.espn.com/mlb/stats/parkfactor/_/year/2018}.
Today, many fans are familiar with these metrics.
In this paper, this PF is called the conventional PF.

The author found some academic papers discussing PF \cite{RePEc:bpj:jqsprt:v:4:y:2008:i:2:n:4}.
This paper discusses the consistency of the conventional PF and concludes that this simple and familiar definition should be modified.

In this paper, I propose a pairwise comparison method to extract the PF from the results.
In particular, every plate appearance (PAs) is considered as a matchup between the batter's team and pitcher's team with a park.
The outcome of every PA (e.g., single, walk, home run, etc.) is explained through a logistic regression model.
Both the proposed and the conventional PFs are evaluated by the log-loss performance index.
The comparison shows the advantage of the proposed PF against the conventional PF.


\section*{Park factor (PF)}
\label{sec:defPF}
\subsection*{Definition of conventional PF}
ESPN's website \cite{http://www.espn.com/mlb/stats/parkfactor/_/year/2018} discloses the definition of the conventional PF.
Let $PF_a^{HR}$ be a PF on home runs of park $a$ where the team $A$ is based.
The definition is
\begin{equation}
\label{eqn:defPF}
PF_a^{HR}=\frac{ \frac{HS_{home}+HA_{home}}{Games_{home}} }
{\frac{HS_{road}+HA_{road} }{Games_{road}} },
\end{equation}
where $HS, HA,$ and $Games$ denote the number of home runs scored, home runs allowed, and games, respectively.
The subscripts $home$ and $road$ means that the numbers are counted for the home and road games.
PF can be similarly defined for the other events that could be expressed by binary variable per PA, i.e., single, double, triple, and walk.
The value $1.0$ is the baseline for a neutral park.
A park with a PF of more than $1.0$ favors the batters, while the one with PF less than $1.0$ favors the pitchers.

Under the definition of (\ref{eqn:defPF}), the log-loss performance index deteriorates for the events of single and walks. This will be discussed later in this paper.
This fact means that the definition of the conventional PF (\ref{eqn:defPF}) can not accurately evaluate the characteristics of ballparks.

\subsection*{Discussions in previous studies}
There is limited literature on PF.
The author found some academic papers including \cite{RePEc:bpj:jqsprt:v:4:y:2008:i:2:n:4}.
(Whether the discussions on websites on SABRMetrics can be considered ``academic'' or not  may be controversial.)
This paper \cite{RePEc:bpj:jqsprt:v:4:y:2008:i:2:n:4} pointed out the inconsistency in the definition of the conventional PF. It especially intends to calculate a robust PF under an unbalanced number of matches.
However, the results of interleague matches were not included.

\section*{Proposed method}
\label{sec:propsoed}
The value of the conventional PF (\ref{eqn:defPF}) can be easily calculated using boxscores.
In simple boxscores, in addition to the runs, the number of home runs and hits are usually recorded.
It is not necessary to check the results of every plate appearance.

Recently, websites such as Retrosheet\cite{https://www.retrosheet.org/} or Baseball Reference\cite{https://www.baseball-reference.com/}, have begun to disclose detailed match results for every plate appearance.
In this paper, a pairwise comparison method based on every plate appearance is proposed.
In particular, assume that each plate appearance can be modeled as a match-up between a batter's team and a pitcher's team plus a ballpark.
The effects of each ballpark are distilled by using a logistic regression method.

\subsection*{Formulation}
Let $i, j\in\{1, \cdots, N_T\}$ and $N_T$ be the indices of teams and the number of teams, respectively.
Let $k\in \{1, \cdots, N_P\}, N_P\geq N_T$ be the index of ballpark.
In MLB, all teams have their own ballpark. Most games are held in these ballparks.
For $i, j, k\in \{1, \cdots, N_T\}$, the index of a team and its ballpark are the same.
$k\in \{N_T+1, \cdots, N_P\}$ denotes the other ballparks.
For instance, MLB has played multiple regular-season games outside of its base ballparks in the USA and Canada \cite{https://www.baseball-reference.com/boxes/OAK/OAK201203280.shtml}.

The team $i$ has two strength parameters for batting (offense) and pitching (defense) which are denoted as  $b_i$ and $d_i$, respectively.
The park $k$ has one parameter $r_k$ that explains PF.
A result of a event (e.g,, home run) at plate appearance $l$ is denoted by $x_l\in \{0,1 \}$.

A plate appearance $l$ is considered as a match-up between a batter's team $i$ and a pitcher's team $j$ with a ballpark $k$.
A probability of the considered event $p_{i,j,k}$ is modeled by the following logistic regression model.
\begin{equation}
\label{eqn:defpijk}
p_{i,j,k}=\frac{1}{1+\exp\left(-\left(b_i - d_j - r_k \right) \right)}.
\end{equation}

The sum of squared error between $p_{i,j,k}$ and $x_l$  is denoted as $J$.
\begin{equation}
J=\sum_{j\in {\mathrm {all~ PA}}}\left(p_{i,j,k}-x_l \right)^2.
\end{equation}

All parameters are updated using steepest descent method.
\begin{equation}
b_i\leftarrow b_i-\alpha \frac{\partial J}{\partial b_i},
\end{equation}
\begin{equation}
d_j\leftarrow d_j-\alpha \frac{\partial J}{\partial d_j},
\end{equation}
\begin{equation}
r_k\leftarrow r_k-\alpha \frac{\partial J}{\partial r_k},
\end{equation}
where $\alpha>0$ is a learning coefficient.

The parameters are calculated for different events.
The event is distinguished by superscript on the parameters.
For instance, $r_k^\mathrm{HR}$ and   $r_k^\mathrm{H}$ are the paremters of the park $k$ of home run and single, respectively.

$r_k$ should be converted to compare with the conventional PF.
\begin{equation}
\label{eqn:proposedPF}
\overline{PF}_k=\frac{\frac{1}{1+\exp\left(-\left(E(b) - E(d) - r_k \right) \right)}}{\frac{1}{1+\exp\left(-\left(E(b) - E(d) - E(r) \right) \right)}},
\end{equation}
where $E(\cdot )$ denotes the average of the associated parameter.
\section*{Calculation and evaluation}
\label{sec:result}

The data of 1550 thousand plate appearances from 2010 to 2017 seasons are collected from Baseball Reference\cite{https://www.baseball-reference.com/}.
The following five PFs are calculated; home runs (HR), single (H), double (2B), triple (3B), and walk.

\begin{figure}[h]
\begin{center}
	\includegraphics[width=0.95\columnwidth ]{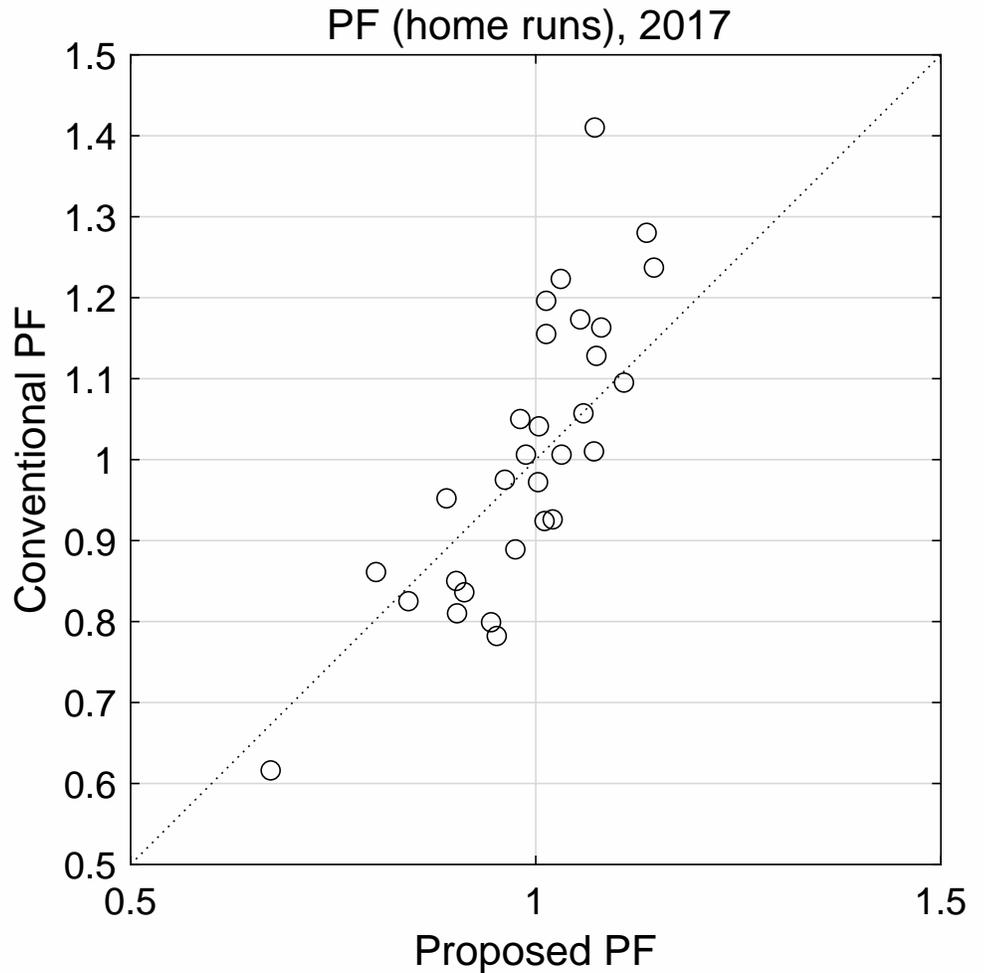}
	\caption{Proposed and conventional PFs in MLB 2017}
	\label{fig:PF_HR_2017}
\end{center}
\end{figure}

Figure \ref{fig:PF_HR_2017} depicts the PF of home runs in the 2017 season.
The horizontal and vertical axes denote the proposed (\ref{eqn:proposedPF}) and the conventional PFs, respectively.

The correlation coefficient is $0.81$. These PFs have a strong correlation.

\subsection*{Performance evaluation}
In this section, both PFs are evaluated.

As a baseline, assume the model in which all parks have the same characteristics.
In this model, the probability of the occurrence of the event is the same for every plate appearance.
For example, in the 2017 season, 6105 home runs were launched in 191195 plate appearances.
The probability of home runs is assumed to be constant, $p_l=0.03193 \equiv p_\mathrm{average}$.

Fitness of the model is evaluated by the log-loss function.
\begin{equation}
LogLoss=E( -x_l\log_2 p_l -(1-x_l)\log_2(1-p_l))
\end{equation}
The log-loss value is $0.20398$ with $p_l=0.03193$.
The baseline values are listed in Table \ref{tab:baseLogLoss}.

\begin{table}[h]
\begin{center}
\caption{Base logarithmic loss}
\label{tab:baseLogLoss}
\begin{tabular}{cccccc}\hline
 & HR & H & 2B & 3B & Walk \\ \hline \hline
2017 & 0.203984  & 0.586163  & 0.249777  & 0.038876  & 0.413728  \\ \hline
2016 & 0.191258  & 0.594936  & 0.245783  & 0.042073  & 0.399136  \\ \hline
2015 & 0.173873  & 0.603291  & 0.244936  & 0.044745  & 0.381862  \\ \hline
2014 & 0.152705  & 0.608606  & 0.242645  & 0.041048  & 0.380149  \\ \hline
2013 & 0.165164  & 0.606779  & 0.244254  & 0.038115  & 0.390933  \\ \hline
2012 & 0.172848  & 0.600476  & 0.245680  & 0.044300  & 0.392314  \\ \hline
2011 & 0.162549  & 0.604597  & 0.247374  & 0.042836  & 0.397113  \\ \hline
2010 & 0.163588  & 0.606679  & 0.248201  & 0.041593  & 0.410027  \\ \hline
\end{tabular}
\end{center}
\end{table}

By the definition (\ref{eqn:defPF}), $p_l$ is estimated py multiplying PF of the park $k$ with $p_\mathrm{average}$.
For instance, the PF of home runs at Coors Field is $1.195$ in the 2017 season.
Therefore, the probability of home runs in this park is calculated as $1.195\times p_\mathrm{average}=0.03816$.
Here, the information on teams could not be used because the conventional definition does not include this data.

The proposed method calculates the probability using (\ref{eqn:proposedPF}).
The proposed model includes not only the park but the offense (batting) and defense (pitching and fielding) strength of each team.
Therefore, the obtained $r_k$ can be separated from the characteristics of the players and can distill the pure effect of the parks.

\begin{figure}[h]
\begin{center}
	\includegraphics[width=0.95\columnwidth ]{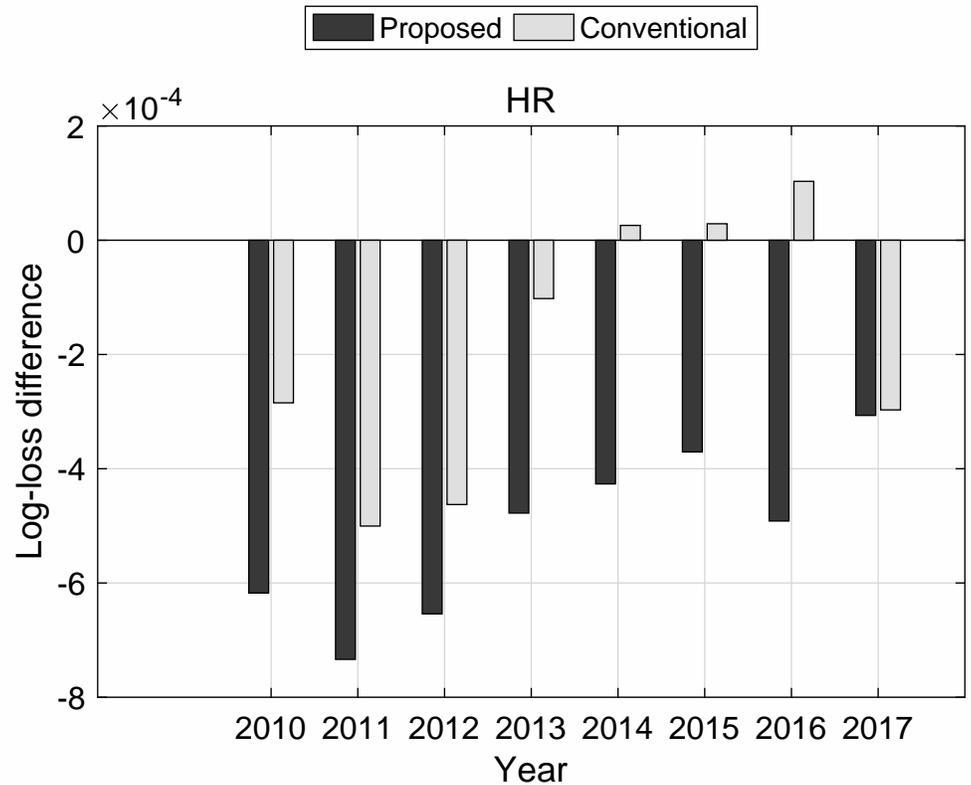}
	\caption{Performance improvement versus the average (HR)}
	\label{fig:performanceHistory_HR}
\end{center}
\end{figure}

\begin{figure}[h]
\begin{center}
	\includegraphics[width=0.95\columnwidth ]{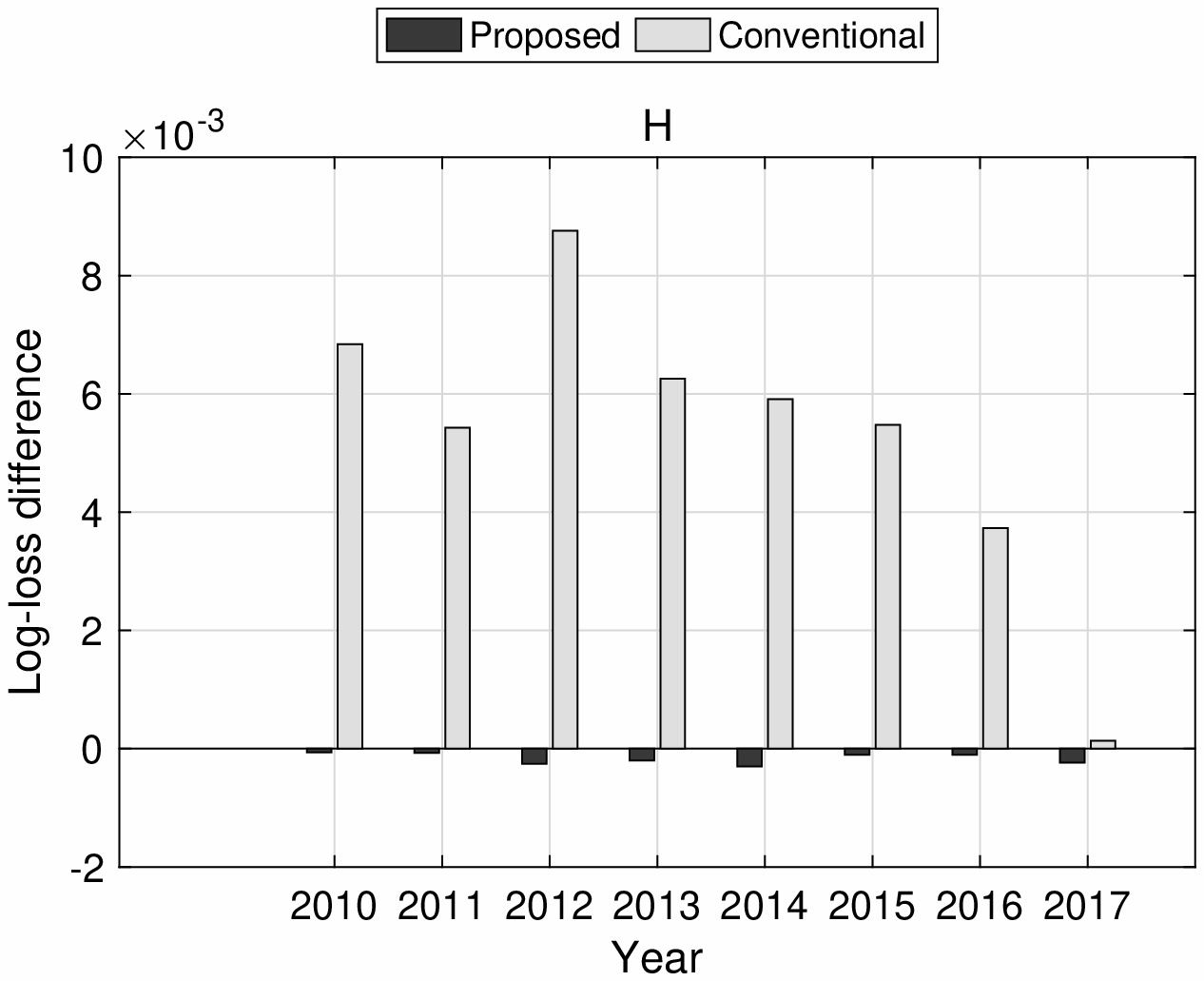}
	\caption{Performance improvement versus the average (H)}
	\label{fig:performanceHistory_H}
\end{center}
\end{figure}

\begin{figure}[h]
\begin{center}
	\includegraphics[width=0.95\columnwidth ]{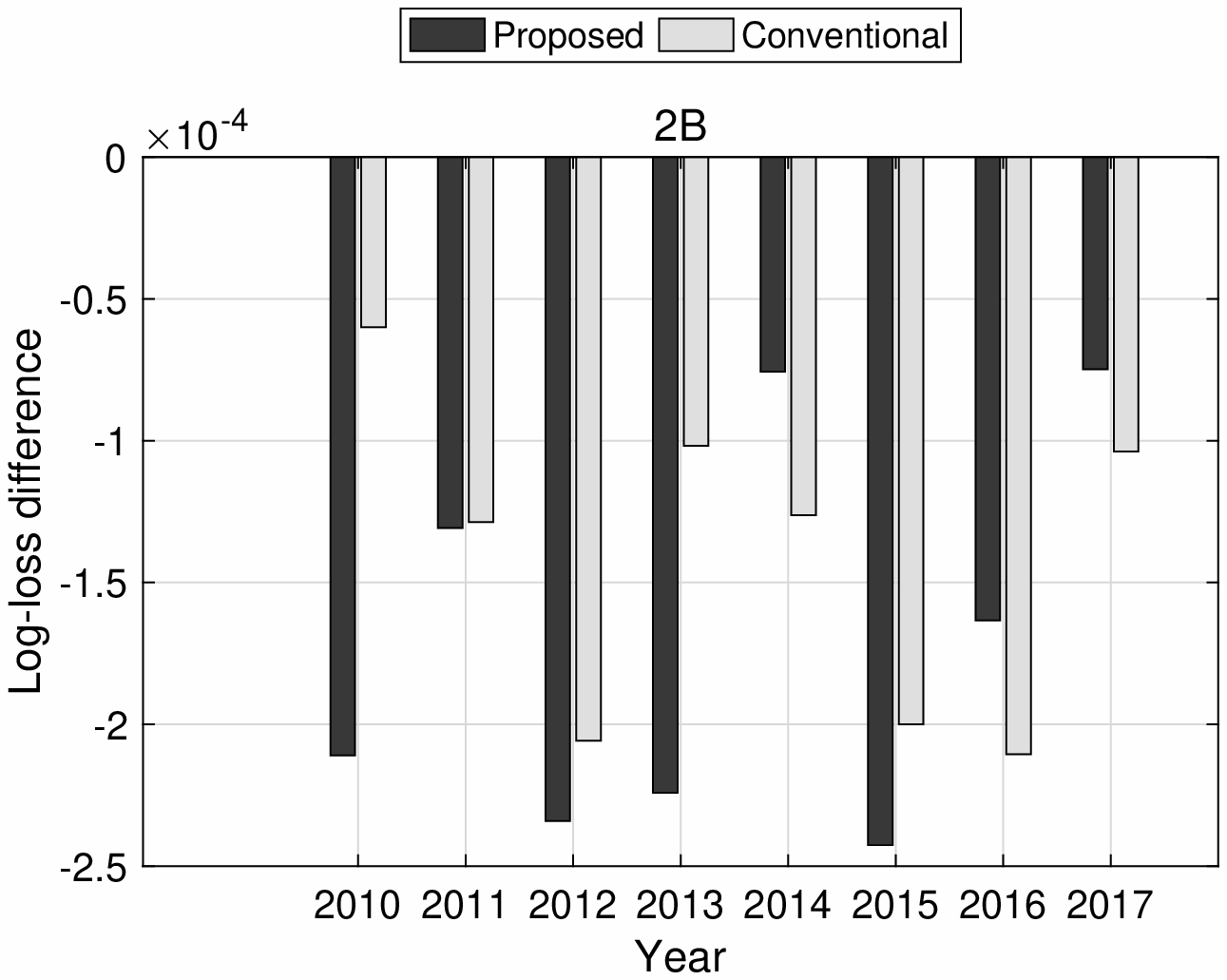}
	\caption{Performance improvement versus the average (2B)}
	\label{fig:performanceHistory_2B}
\end{center}
\end{figure}

\begin{figure}[h]
\begin{center}
	\includegraphics[width=0.95\columnwidth ]{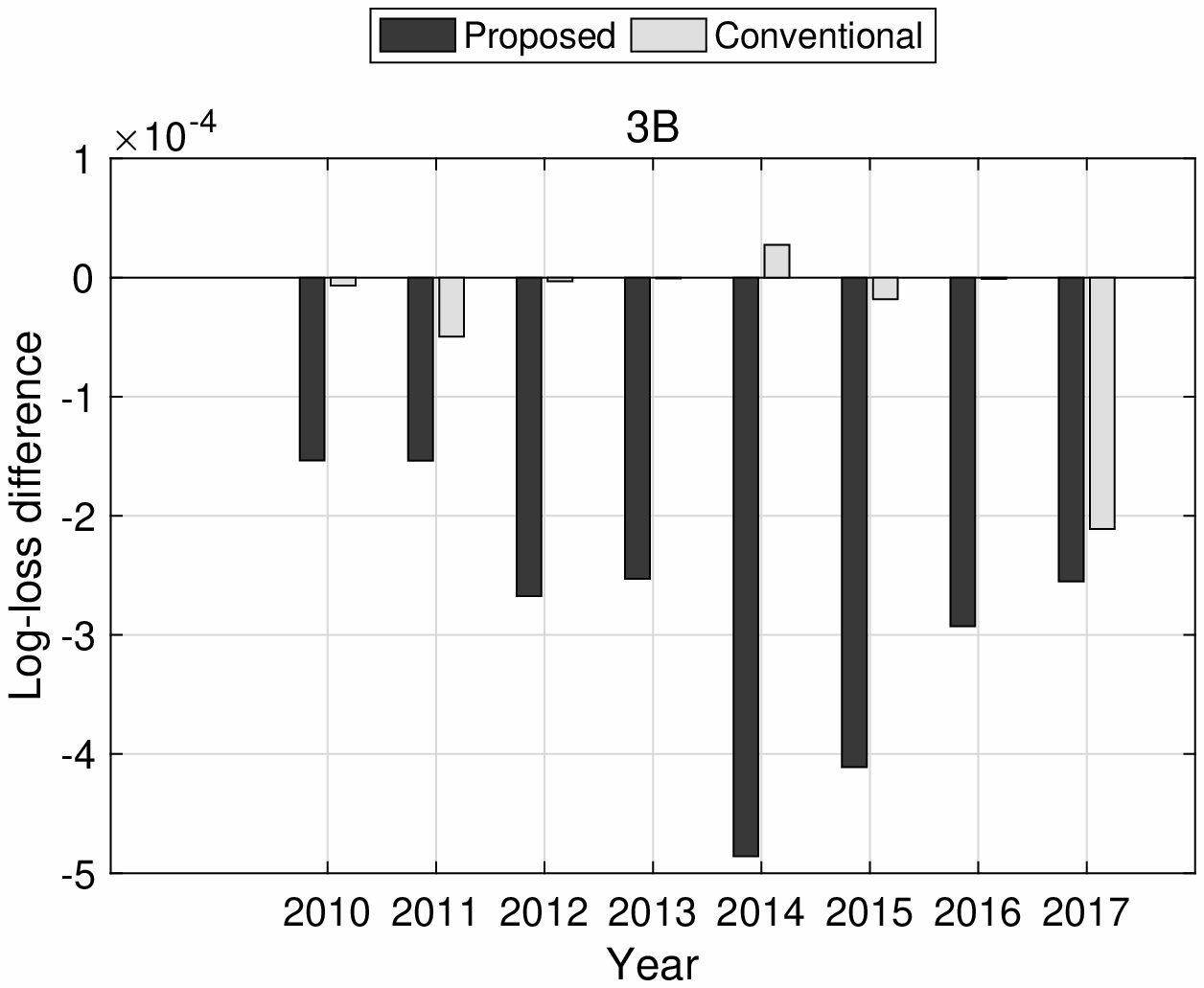}
	\caption{Performance improvement versus the average (3B)}
	\label{fig:performanceHistory_3B}
\end{center}
\end{figure}

\begin{figure}[h]
\begin{center}
	\includegraphics[width=0.95\columnwidth ]{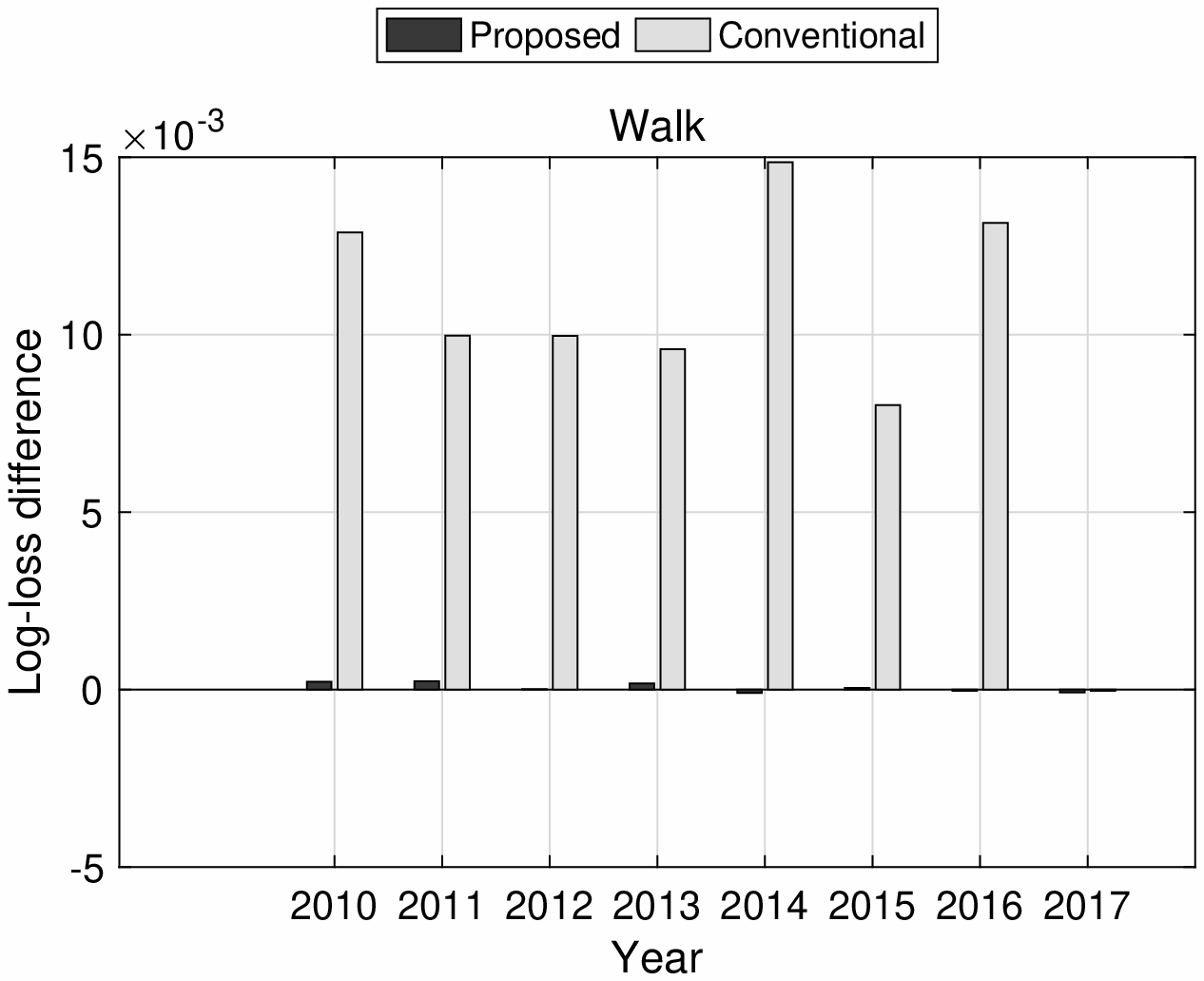}
	\caption{Performance improvement versus the average (Walk)}
	\label{fig:performanceHistory_Walk}
\end{center}
\end{figure}

The figures \ref{fig:performanceHistory_HR}--\ref{fig:performanceHistory_Walk} illustrates the log-loss difference between the baseline.
Note that the negative difference implies a performance improvement.


\subsection*{Discussion}

The conventional PFs can improve the fitness from the baseline for long hits (2B, 3B, and HR).
However, the fitness is deteriorated for single and walk.
By intuition, a walk is the result of a match-up between a batter and a pitcher (or a battery).
It is not a natural hypothesis that the size of the ballpark has a stronger effect than the players on the probability of walk.
This result shows that the conventional PF (\ref{eqn:defPF}) should be modified.

The proposed PF improves the fitness of the conventional PF for HR, 3B, and H.
The improvements are similar in 2B among the two PFs.
Note that the proposed method indicates that the park factor for single and walk are negligible.
The probability of single and walk should not be explained by the park effect.
This result demonstrates that the proposed PF can accurately evaluate the effect of the ballparks.

\section*{Conclusion}
\label{sec:conclusion}

This paper considers the PF that quantifies every ballpark's characteristics in baseball.
In this study, I assume that each plate appearance can be modeled as a match-up between a batter's team and a pitcher's team plus a ballpark.
The effects of each are distilled by using a logistic regression method.

The verification was based on the results of more than 1.5 million plate appearances from the 2010 to 2017 MLB seasons.
Numerical verification showed that the conventional PF could not improve the fitness for single and walk.
The proposed method could improve fitness compared with the baseline value more than the conventional one.
Moreover, a natural insight -- the size of the park does not affect the probability of walks -- could be drawn by the proposed method.

\end{document}